\def \cm{~\rm{cm}}
\def \s{~\rm{s}}
\def \km{~\rm{km}}

\def \K{~\rm{K}}

\def \erg{~\rm{erg}}

\def \yr{~\rm{yr}}
\def \Myr{~\rm{Myr}}

\def \kpc{~\rm{kpc}}

\def \muG{~\rm{\mu~{\rm G}}}
\documentclass[preprint]{aastex}
\shortauthors{Soker}
\begin{document}

\title{MAGNETIC FIELDS IN COOLING FLOW CLUSTERS: A CRITICAL VIEW}

\author{Noam Soker\altaffilmark{1}}

\altaffiltext{1}{Dept. of Physics, Technion, Haifa 32000, Israel;
soker@physics.technion.ac.il}

\begin{abstract}
Shortly after the first results of \emph{Chandra}  and \emph{XMM-Newton} appeared,
many researchers in the field abandoned the
term ``cooling flow clusters'' in favor of the name ``cool core clusters''.
This change, I argue, has been causing damage by promoting the view
that there is no substantial cooling in these clusters.
In this contribution I discuss the following points, with emphasize on the last one that
deals with magnetic fields in cooling flow clusters.
(1) Both AGN-feedback and hot-gas cooling to form stars occur during galaxy formation as well as in
cooling flow clusters.
Ignoring cooling of the intra-cluster medium, as implied by the term ``cool core'', does not encourage
comparative study of AGN feedback in cooling flow clusters with that of galaxy formation.
(2) The line of thought that there is no cooling might lead to wrong questions and research directions.
(3) A key question in both cooling flow clusters and during galaxy formation
is the mode of accretion by the super massive black hole (SMBH).
When cooling is neglected only accretion from the hot phase remains.
 Accretion from the hot phase, such as the Bondi accretion, suffers
from some severe problems.
(4) When it is accepted that moderate quantities of gas are cooling, it becomes clear
that global heat conduction must be substantially suppressed.
This does not favor a globally ordered magnetic field.
As well, it makes global heat conduction unattractive.
\end{abstract}

%{\it Subject headings:}
%{\bf Key words:}
% =====================================================
\section{INTRODUCTION}
\label{sec:intro}
% =====================================================
% =========================
\subsection{Preface}
\label{sec:preface}
% =========================
The field of cooling flow (CF) in galaxies and in clusters of galaxies evolved tremendously
in the last decade (e.g., compare the old reviews by Sarazin 1986 and Fabian 1994 with
those of Peterson \& Fabian 2006 and McNamara \& Nulsen 2007).
It seems as if in some parts the new observations from the two X-ray space observatories
have been coming at a too high rate, in the sense that the community took new
directions before fully digest the new results.
I would go as far as saying that the ``revolution'' was sociological rather than an
astrophysical (scientific) one.
Early observations from the then new X-ray observatories (e.g. Kaastra et al. 2001;
Peterson et al. 2003) {\it put an upper limit} on the cooling rate.
As the typical upper limit on cooling rates was much below the value expected from
first-generation cooling flow models (Sarazin 1986; Fabian 1994), many took it as
if there is no CF at all.
The term ``cool core clusters'' was born, replacing in many papers the term ``cooling flow clusters''.
This change in the name has been causing some damage by promoting the view that there is
no substantial cooling in these clusters.

In the present paper I will focus on, and critically examine, the state of three research
areas that involve magnetic fields as major or minor ingredients.
In section \ref{sec:conduction} I review the role of heat conduction.
The results of XMM-Newton and Chandra renewed interest in the role of heat conduction.
Heat conduction was considered in the context of first-generation CF models.
In particular, Bregman \& David (1988; hereafter BD88) conducted a thorough study
and showed that practically, heat conduction cannot suppress CFs.
This work was ignored by too many papers in recent years.
In section \ref{sec:morpology} I critically examine recent papers that study
large scale coherence magnetic fields.
In section \ref{sec:bubbles} I argue that there is no need for attributing
the stability of X-ray deficient bubbles to magnetic fields.
As for a summary, section \ref{sec:damage} below can serve as the summary, as well as the
main motivation for this paper.
This is not a review paper, and I will cite only a few examples in each topic.
But first let me elaborate on the change of terms from
'`cooling flow clusters'' to ``cool cores clusters''.

% =========================
\subsection{Change of names: damage control}
\label{sec:damage}
% =========================
Interestingly, the planetary nebulae (PNs) community did not change the name even after
PNs have been shown not to be planets. Neither the brown dwarf community changed the name
after the spectra of brown dwarfs were found not to be brown.
The change in names from CF to cool-core was part of a trend to start things all over
again, unjustifiably ignoring previous theoretical results (and observations), in particular
those that were obtained with the first-generation CF models in mind.

Not before long observations revealed the expected signatures of CFs, but at lower
cooling rates. The CF signatures include star formation (see summary of results by
Rafferty et al. 2008 and a recent paper with more references by O'Dea et al. 2010),
molecular gas (e.g., Wilman et al. 2009),
and dust (e.g., Edge et al. 2010).
The CF model where the cooling rate is much below the one expected when no heating
is present is termed the \emph{moderate cooling flow model.}
It was suggested in Soker et al. (2001), and was developed in a series of papers since then.

When one takes it to the extreme that the basic state is a ``cool-core'',
and that there is no CF at all, unnecessary questions might arise.
Such is the question possed by Cattaneo et al. (2009) of
`` . . .why real clusters depart from an 'ideal' feedback loop that is $100\%$
efficient in suppressing cooling and star formation.''
In the moderate CF model the natural state is of cooling, a moderate one.
That the heating mechanism cannot prevent cooling is essential for cold
clumps to feed the central super massive black hole (SMBH) and lead to an AGN activity.
This mode of feeding is the cold feedback mechanism (Pizzolato \& Soker 2005, 2010),
that is a key ingredient in the moderate CF model.
In addition, when no cooling is allowed in the model, the accretion mode to the SMBH must
be from the hot intracluster medium (ICM) phase. Such is the Bondi accretion.
However, the Bondi accretion has severe problems in explaining feedback in
CF clusters (Soker et al. 2010; McNamara et al. 2010).

I find the damage in the change of names to be more than semantical.
\begin{enumerate}
\item  {\it Comparison with feedback during galaxy formation.}
During the process of galaxy formation AGN feedback is likely to occur. In addition,
large quantities of the interstellar medium (ISM) are cooling to form stars.
Ignoring cooling of the ICM (such as implied by the term ``cool core''), does not
encourage comparative study of AGN feedback in CF clusters with that of galaxy formation.
Such a comparison is natural when moderate cooling in CF clusters is considered (Soker 2010).
\item {\it Looking under the wrong lamp}. As discussed above, the line of though that there
is no cooling might lead to wrong questions and research directions.
\item {\it The mode of feedback.} A key question in both CF clusters and during galaxy formation
is the mode of accretion by the SMBH. When one neglects cooling, then only accretion from the
hot phase remains. Accretion from the hot phase, such as the Bondi accretion, suffers
from some severe problems (Soker et al. 2010; McNamara et al. 2010).
\item {\it Magnetic field morphology.} When it is accepted that moderate quantities of gas
are cooling, it becomes clear that global heat conduction must be substantially suppressed.
This does not favor a globally ordered magnetic field (see section \ref{sec:morpology}).
This also makes heat conduction unattractive (section \ref{sec:conduction}).
This is the topic of the present paper.
\end{enumerate}

It is true that the change in names by itself does not imply any of the above.
However, judging from the literature, this trend has been happening too many times
in the last decade.

% =====================================================
\section{THE FAILURE OF HEATING BY GLOBAL HEAT CONDUCTION}
\label{sec:conduction}
% =====================================================

In this section I argue that heat conduction cannot work by itself to explain
properties of CF clusters.
An extra heating source is required in such models, and heat conduction becomes a
bystander.
Further more, if instead of a ``cool core'' one is aiming at building models for
more realistic moderate CF clusters, then the heat conduction is not even a bystander.
It is simply not required.

% =========================
\subsection{Generic behavior of heat conduction}
\label{sec:behavior}
% =========================

Heat conduction in CF clusters, as well as its problematic nature,
was discussed before the work of BD88.
However, BD88 conducted a thorough study that basically holds to present.
BD88 finds that either heat conduction is unstable altogether, or that under
some circumstances a fine tuning of the parameters is required.
The fine tuning problem was noted already by Binney \& Cowie (1981).
Such a fine tuning will not operate in the observed non-homogeneous ICM of CF clusters.

With a very simple model (Soker 2003) I strengthened the results of BD88 by showing
 that models based on heat conduction alone are unstable.
Perturbations extending over a large fraction of the CF region
and with an amplitude of $\sim 10 \%$, will grow to the non-linear regime within a
Hubble time.
This result was further strengthened with a more sophisticated study conducted
by Zakamska \& Narayan (2003; {{{ see also Kim \& Narayan 2003). }}}

This is a generic behavior of heat conduction: it is unstable to global modes.
A configuration based on heat conduction cannot hold for the age of the cluster,
and an additional heating source must be added.
Because heat conduction is less efficient in the inner regions (as they are cooler),
exactly where radiative cooling rate is very high (as it is denser there),
the extra heating source must be very efficient in the inner regions.
Basically, the extra heating source becomes the main heating source.
Over all, global heat conduction fails to account for the properties of CF clusters,
and at most plays a minor role.
When AGN activity is considered, and allowance for some cooling is added (as in the moderate
CF model), it is unavoidable to conclude that global heat conduction should be abandoned
altogether (section \ref{sec:morpology}).

Basically, heat conduction cannot work by itself. If an extra heating source is added, then
there is no need any more for the heat conduction.
Let me give a number of examples.

% =========================
\subsection{Adding an inner heating source: back to a moderate CF model}
\label{sec:add1}
% =========================

Ruszkowski \& Begelman (2002) claim to stabilize the heat conduction process by adding AGN
feedback heating in the inner region of their 1D spherical model.
The AGN basically heats the inner region.
In their studied case the accretion rate comes from the inner region
of the computational zone.
The accretion rate there is $1.76 M_\odot \yr^{-1}$.
As they themselves write, cooling of gas by local thermal instabilities
will lead to an even higher total cooling rate.
Their model is actually a moderate CF model, with conduction playing some
adjusting role.
They still have the problem that the accretion comes from the inner region, and the inflow
time from $r \sim 1 \kpc$ might be longer than the cooling time there (this
time delay is not considered by them). This is one of the problems encountered by
the Bondi accretion process (Soker et al. 2010).

 I suggest that incorporating a different heating scheme by the AGN, and
including the cold feedback accretion, would remove the need to include heat conduction.
More than that, in a recent paper Guo et al. (2010) found that for stability, the efficiency
of transferring mass accreted from the inner computational zone to AGN power,
$L_{\rm AGN} = \epsilon \dot M_{\rm BH} c^2$, should be in some cases as large as
$\epsilon = 0.3$.
Such a high efficiency clearly can work only with more mass accreted from
extended regions (if one want to save such a model).
To summarize, to save such a model one must turn to consider a moderate CF model with a cold feedback
mechanism and AGN feedback. Heat conduction does not really help.

% =========================
\subsection{Adding turbulence: back to a moderate CF model}
\label{sec:add2}
% =========================
Let me take a recent example, and critically review the results of Parrish et al. (2010).
Parrish et al. (2010) consider heat conduction as the heating source. They add to their
simulations a turbulence, and follow the geometry of the magnetic fields, and from that
calculate heat conduction.
They find that whether the inner region catastrophically cools or is heated up depends
on their initial added turbulence.
This is a generic behavior of the instability attached to heat conduction, heating or
catastrophic cooling (Soker 2003; Zakamska \& Narayan 2003; {{{ Kim \& Narayan 2003). }}}

Consider energetic in their models that heat up.
They claim that the major heating is by the heat conduction, while heating by
turbulence is negligible.
Instead, it seems as if the entire inner region is heated by two other processes:
mixing of hotter outer regions by the turbulence, and dissipation of the turbulence.
This is clearly seen in the inner 20kpc of their hot solutions, where the temperature
gradient is flat, or even slightly negative.
Namely, this region, which is most prone to thermal instabilities
(because of its higher density and lower temperature) is not heated by heat conduction.

In their cooling-catastrophe solutions they claim that heat conduction is inhibited by
the heat-flux-driven buoyancy instability (HBI).
However, from their figure 2 it is seen that the reduced conduction occurs only
in the inner regions of $r<20 \kpc$.
Yet, the cooling catastrophe occurs for $r < 40 \kpc$.
In any case, even in the hot solutions the heat conduction does not play a role inside
$r<20 \kpc$.
The same holds for similar calculations performed by Ruszkowski \& Oh (2010):
there is a negative temperature gradient in the inner $20 \kpc$, and a flat one inner to
$30 \kpc$. Namely, heat conduction does not heat the inner region. It is mixing from
outer regions, or turbulent dissipation (see also David \& Nulsen 2008).

I note that Kunz et al. (2010) assume the presence of turbulence to heat the ICM.
They do not fully solve the energy problem, as they assume that there is
enough turbulent energy to offset cooling, but do not discuss the source of the
turbulent energy, and how feedback is maintained with the energy in the turbulence itself.
They basically suggest a mechanism to dissipate turbulent energy, which might indeed operate
in the ICM.
Kunz et al. (2010) assume small scale magnetic fields, hence their model does not suffer
from problems associated with the assumption of larger scale ordered magnetic fields
(as some other models that are based on heat conduction do suffer from).

To summarize my view on the results of Parrish et al. (2010),
I would say that with a complicated 3D numerical code they have reobtained the thermal instability
of models based on global heat conduction (BD88; Soker 2003; Zakamska \& Narayan 2003; {{{ Kim \& Narayan 2003). }}}
In addition, a careful examination of their results in the inner
$\sim 20 \kpc$ (as well as of Ruszkowski \& Oh 2010) show that another heating source
is presence there.
This source is probably dissipation of turbulenct energy, or mixing of outer
hotter regions (it is impossible to tell from their paper).
This sends us back to section \ref{sec:add1}, and to a moderate CF model.

{{{ After the pre-meeting version of this paper has been published,
another demonstration that heat conduction cannot work was successfully performed.
Ruszkowski \& Oh (2011) built a model where turbulent heat diffusion is important
in the cluster center, while conduction is important in the outer regions.
Evolving their initial model for several Gyr, they either get catastrophic cooling,
or they heat the central region to be only marginally cooler than the
outer region. Near the center there is a negative temperature gradient, contrary to observations.
They definitely don't obtain a ``cole core'', but rather show the
unstable nature of heat conduction with turbulent heat diffusion:
Either heating to remove the ``cole core'', or catastrophic cooling occur, but not
the observed temperature profile in CF clusters. }}}

% =====================================================
\section{ON THE MAGNETIC FIELD MORPHOLOGY}
\label{sec:morpology}
% =====================================================

In this section I argue that the ICM in CF clusters is perturbed by AGN activity
to a degree that makes any process that is based on global morphology of magnetic
fields, such as the heat buoyancy instability (HBI), useless.
Moreover, large scale magnetic fields will inhibit local thermal instabilities
that are required by observations.

% =========================
\subsection{General considerations for heat conduction}
\label{sec:field}
% =========================

\begin{figure}
\begin{center}
\vskip11mm
\includegraphics[scale=0.6]{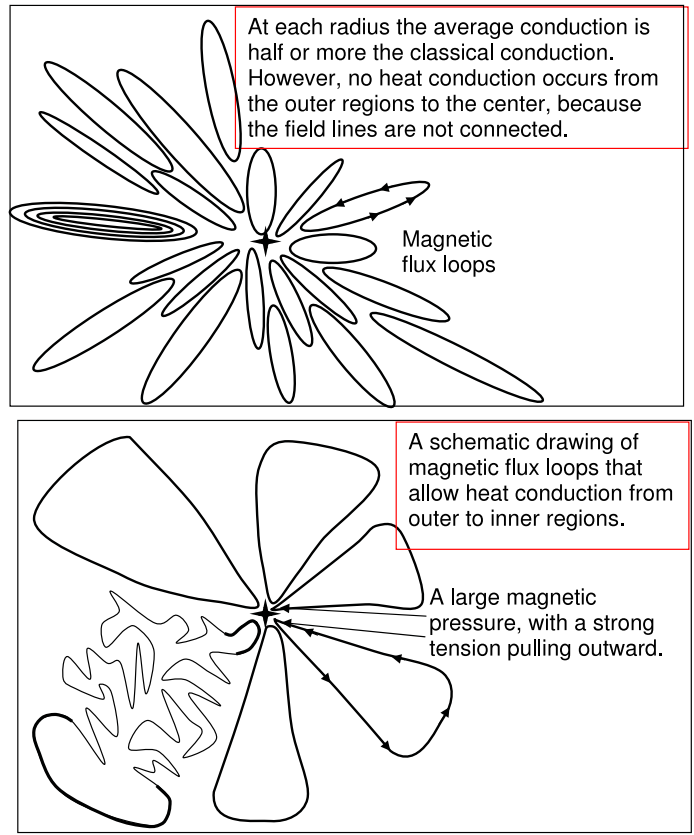}
\end{center}
\caption{\emph{Upper panel:}
A schematic magnetic field structure composed of magnetic flux loops.
Magnetic fields exist inside the loops, as drawn for clarity only for the left loop.
Such a structure has an average conduction flux at each radius which is half or
more of the classical value (no magnetic fields). However, there is no heat conduction from
outer to inner regions.
\emph{ Lower panel:}
A schematic drawing of a magnetic structure that allows heat conduction
from outer to inner regions. The magnetic field becomes extremely strong near the center.
It will reconnect and the magnetic tension will pull it outward (Zoabi et al. 1996).
A tangled magnetic field (as depicted in the lower left), does not overcome this problem
because its ends must still reside in the very outer and very inner regions.
}
\label{fig:fig1}
\end{figure}

In some papers (e.g., Parrish \& Quataert 2008; Bogdanovic et al. 2009)
the heat conduction is calculated by taking the average heat flux at a specific radius
according to
\begin{equation}
\overrightarrow Q=Q_0 \frac{ \overrightarrow B ~ \overrightarrow {B} \cdot \overrightarrow \nabla T}
{B^2 \vert \overrightarrow \nabla T \vert},
\label{eq:cond}
\end{equation}
where $Q_0$ is the conduction in a non-magnetized plasma with the same properties,
$\overrightarrow{B}$ is the magnetic field, and $T$ is the temperature.
In the upper panel of Figure \ref{fig:fig1} a schematic magnetic field topology is drawn.
Each loop represents a magnetic loop. Namely, there are magnetic fields inside the loops as well,
as drawn only for the loop at the far left. At each cluster radius $r$, the average heat
conduction is $\bar Q \ga 0.5 \bar Q_0$, as at least half the field lines are radial.
However, it is clear that there is no connection between the outer and inner regions,
as the typical length of the flux loops is shorter than the size of the cooling region.
% FFFFFFFFFFFFFFFFFFFFFFFFFFFFF
% FFFFFFFFFFFFFFFFFFFFFFFFFFFFF

To conduct from the outer regions to the inner regions the magnetic fields lines must connect the
inner and outer regions (for reconnection of magnetic flux loops see Norman \& Meiksin 1996).
A schematic drawing is in the lower panel of Figure \ref{fig:fig1}.
The loops are stretched all the distance from the outer to the inner regions.
The magnetic field lines can be tangled, as depicted by the thin line of the
lower-left loop.
However, the very outer and very inner ends of the loop, depicted by thick lines on
the lower-left loop, must reside in the outer and inner regions, and so look the same as
the ends of the schematic loops.
Conservation of magnetic flux implies that the radial component of the magnetic field
behaves as $B_r \propto r^{-2}$, and the magnetic pressure as $P_B \propto r^{-4}$
(Soker \& Sarazin 1990).

Let me use typical values following Soker \& Sarazin (1990).
Taking a magnetic field of $\sim 1 \muG$ at $r=30 \kpc$ and a thermal pressure of
$\sim 3 \times 10^{-10} \erg \cm^{-3}$ there, gives for the magnetic to thermal pressure
ratio $P_B/P_{\rm th} \simeq 10^{-4}$.
For the heat conduction to prevent cooling in the inner regions,
the magnetic field should connect the outer regions with the inner regions.
The magnetic pressure due to the radial component comprise a third of the magnetic pressure.
I find that the magnetic pressure at, say, $r=1 \kpc$, is $\sim (1/3)(30)^4 \simeq 3 \times 10^5$
times that at $30  \kpc$. The thermal pressure increases by a factor of $\sim 3$
(e.g., Blanton et al. 2009 for A2052;  McNamara et al. 2000 for hydra A).
The conclusion is that the magnetic pressure becomes comparable to, or even larger than, the
thermal pressure.
At that stage magnetic field reconnection will change the structure (Soker \& Sarazin 1990),
and magnetic stress will pull the inner parts of the loop outward.

 If the magnetic field lines crosses the center (as in fig.1-left of Bogdanovic et al. 2009),
then they are connected to a limited volume in the outer region.
In this case, more likely the inner region will {\it cool} the outer one, rather than be
heated.

Over all, it seems that there is no way to build a self-consistent magnetic field structure
that will allow global heat conduction to prevent cooling in CF clusters.

% =========================
\subsection{The ICM motion in CF clusters}
\label{sec:velocity}
% =========================

In recent years some papers studied the behavior of magnetic fields in the ICM.
A steady state medium was assumed in many cases, while other introduce turbulence.
Under ideal conditions the heat-flux-driven buoyancy instability (HBI) might develop
(Quataert 2008; Parrish \& Quataert 2008; Bogdanovic et al. 2009; Parrish et al. 2009).
The HBI is a convective, buoyancy-driven instability
that tends to rearrange the magnetic field to be preferentially
perpendicular to the radial temperature gradient, when the temperature gradient is opposite
to that of gravity (Quataert 2008; Parrish \& Quataert 2008).
In the calculations of Parrish et al. (2009) and Bogdanovic et al. (2009)
the typical growth time is ${\rm few} \times 10^8 \yr$ up to $\sim 10^9 \yr$.
This time scale is comparable to, or longer than typical time period between major AGN outbursts
in CF clusters.
These studies start from an already globally ordered magnetic field, and examine the time
it takes for the field to significantly change its global topology.

\begin{figure}
\begin{center}
% \vskip11mm
\vskip -11mm
\includegraphics[scale=0.6]{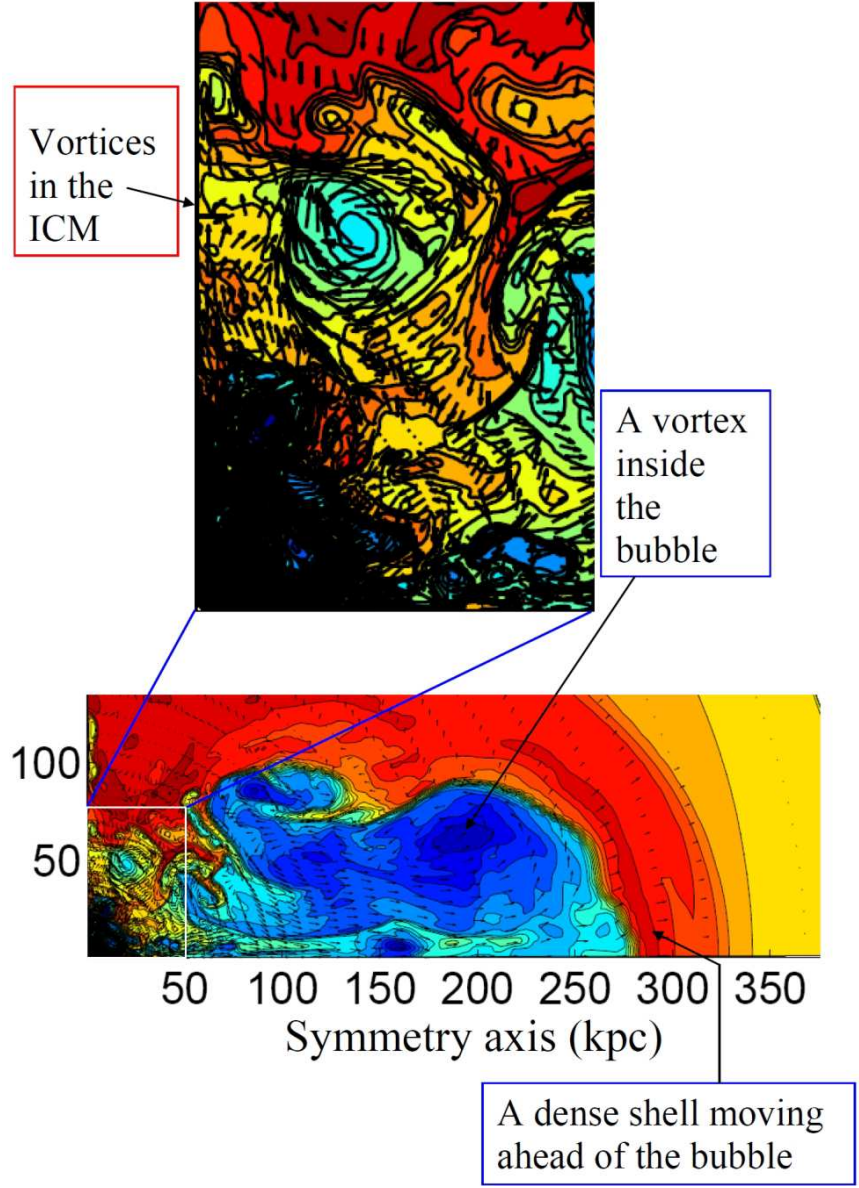}
\end{center}
\vskip -7mm
\caption{ Log-density maps of the evolution of a model of jet-inflated bubbles at
$t=115 \Myr$ (model I from Sternberg \& Soker 2009b).
The units on the axes are kpc. The symmetry (jets) axis is the horizontal axis.
The jet was active for $100 \Myr$ and was turned off at $t=100 \Myr$.
Blue and red are for low and high densities, respectively (see Sternberg \& Soker 2009b for details).
The arrows represent the flow velocity, and are grouped into four velocity bins:
$(i)$ $0.1c_s \leq v_j \leq c_s$ (shortest arrows);
$(ii)$ $c_s \leq v_j \leq7.5c_s$;
$(iii)$ $7.5c_s \leq v_j \leq 15c_s$;
$(iv)$ $15c_s \leq v_j \leq 30c_s$ (longest arrows),
where $c_s=1152 \km \s^{-1}$ is the unperturbed ICM sound speed.
Large vortices that cover the entire inner region will destroy any large scale
structure of the magnetic field. }
\label{fig:fig2}
\end{figure}

The physical mechanism of the HBI is very nice and elegant.
However, I think it is not relevant at all to CF clusters.
The frequently-launched jets, that also inflate bubbles, completely disturb
large regions, even perpendicular to the jets' axis.
This is demonstrated in Figure \ref{fig:fig2}, taken from
Sternberg \& Soker (2009b, their fig. 2).
The simulations were intended to reproduce the structure of the bubbles in the
CF cluster MS0735+7421.
It is evident that large vortices exist perpendicular to the symmetry axis
(the horizontal axis in the figure).
The simulation presented here is a 2.5 dimensional flow.
Namely, the 3D flow has a cylindrical symmetry, and the code includes the $(z, r)$ coordinates only.
In full 3D simulations with higher resolution, the inner region is expected to be more disturbed
even (more vortices on other scales).
The disturbed nature of the ICM in regions perpendicular to the jets' axis
when bubbles are inflated can be seen also in Sternberg et al. (2007), Sternberg \& Soker (2008a),
Falceta-Goncalves et al. (2010), and Morsony et al. (2010).
% FFFFFFFFFFFFFFFFFFFFFFFFFFFFF
% FFFFFFFFFFFFFFFFFFFFFFFFFFFFF

\begin{figure}
\begin{center}
% \vskip11mm
\vskip -11mm
\includegraphics[scale=0.6]{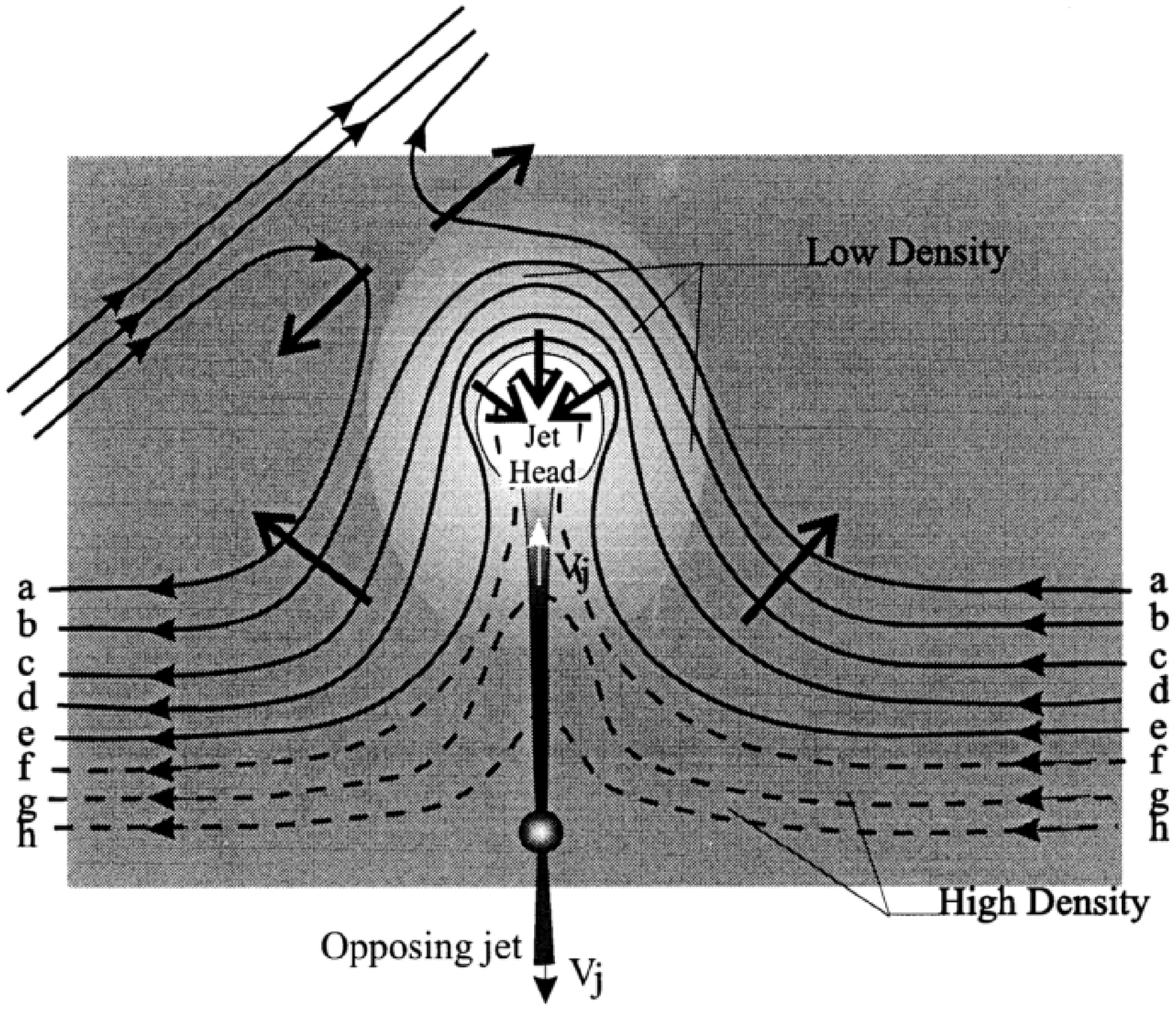}
\end{center}
\vskip -7mm
\caption{Schematic presentation of a radio jet stretching magnetic field lines (from Soker 1997).
The stretching of magnetic field lines followed by reconnection (see also O'Neill \& Jones 2010),
are more important that the self assembly of magnetic fields in CF clusters.
Thick arrows represent magnetic stress; solid lines represent magnetic field lines in the symmetry plane;
dashed lines are magnetic field lines above (or below) the symmetry plane.
The field line marked ``a'' represents reconnection with magnetic field lines farther out in the
cluster which are not stretched by the jet. The field line marked ``f'' represents field lines
just before reconnecting behind the jet's head.    }
\label{fig:fig3}
\end{figure}

In addition to the flow imposed by inflated bubbles, narrow jets
that do not inflate bubbles can also alter the structure of
ordered magnetic field lines. This was already noticed in the
frame of first generation CF models (Soker 1997), and is
schematically drawn in \ref{fig:fig3}. In that paper the
stretching and reconnection of magnetic field lines were
discussed. A similar process of magnetic field lines stretching
was recently studied with 3D magnetohydrodynamic simulations by O'Neill \& Jones (2010) and
Huarte-Espinosa et al. (2010).
Huarte-Espinosa et al. (2010) used the compressed and stretched field lines to calculate the expected
rotation measure.
O'Neill \& Jones (2010) further point out the complicated structure that is formed by stretching, twisting, and
reconnection, of magnetic field lines.
A similar process of magnetic field lines stretching can occur when a galaxy moves
through the ICM (Pfrommer \& Dursi 2010, who termed the process magnetic draping).
% FFFFFFFFFFFFFFFFFFFFFFFFFFFFF
% FFFFFFFFFFFFFFFFFFFFFFFFFFFFF

I note here that in a recent paper Million et al. (2010) find that the diffuse cluster
gas at a given radius in M87 (the virgo cluster) is strikingly isothermal.
They further write that
``This suggests either that the ambient cluster gas, beyond the arms, remains relatively undisturbed by AGN
uplift, or that conduction in the intracluster medium (ICM) is efficient along azimuthal
directions, as expected under action of the heat-flux driven buoyancy instability (HBI).''
The HBI explanation is problematic, as Million et al. (2010) themselves notice, their figure 5 shows
that the metals in the ambient ICM are significantly clumped.
This suggest that different clumps had different history.
Why then the magnetic field topology be smooth as required by the HBI explanation?
Instead, the isothermal structure along constant radii can be explained in
the frame of the cold feed back mechanism, as of course is the clumpy metal distribution.
Clumps cooler (hotter) than their surrounding sink (float), and are
remove from their surroundings.

The flow structure obtained in numerical simulations (turbulence,
vortices, bubbles, expanding jets) will inhibit any process by
which the magnetic field orders itself, such as in the HBI.
Instead, magnetic field will be more likely arranges in flux
loops. This is discussed in the next section.

% =========================
\subsection{A more realistic magnetic field morphology}
\label{sec:more1}
% =========================

Some of the results on magnetic fields in CF clusters that were obtained in the frame of the
first-generation CF models are not relevant any more. Such are some parts
in the work of Soker \& Sarazin (1990) that were based on huge mass inflow rates,
and ignored bubbles and jets activity.
On the other hand, some results are relevant to the new generation models as well.
I will briefly mention those that deal with small scale magnetic flux loops (tubes).

The main theme I would like to present here is that the magnetic field lines
are closed on themselves on scales much smaller than the relevant length scale
of $\sim 10-30 \kpc$ in clusters.
These closed structures are termed magnetic flux loops,
and are drawn schematically in Figure \ref{fig:fig1}.

% =========
\subsubsection{Magnetic flux loops}
\label{subsubsec:loops}
\begin{figure}
\begin{center}
\vskip11mm
\includegraphics[scale=0.6]{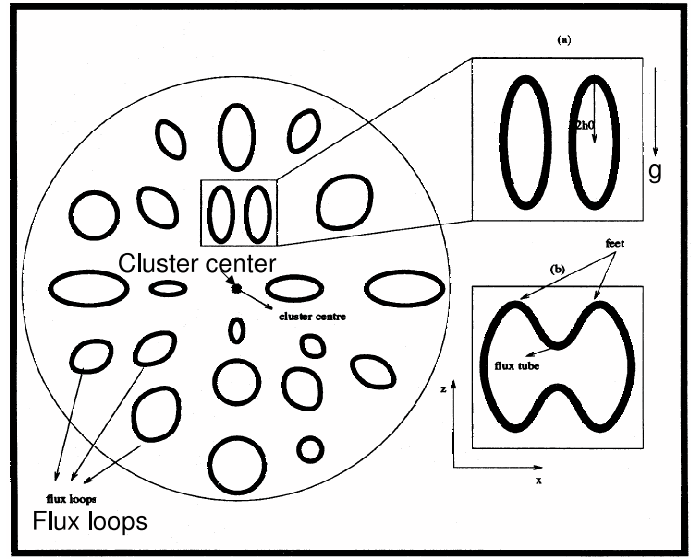}
\end{center}
\caption{A schematic drawing of the magnetic field morphology I propose to
hold in cooling flow clusters (from Zoabi et al. 1996).
In reality the structure is more complicated, with some flux loops taking
a less regular shape as the lower-left flux loop in Figure \ref{fig:fig1}.
Zoabi et al. discussed the possibility that two closed loops reconnect as
depicted in panels a and b. }
\label{fig:fig4}
\end{figure}

% =========
The presence of cold gas in CF clusters (e.g., Salom{\'e} et al. 2008; Wilman et al. 2009;
Edge et al. 2010) suggests that hot gas cools to form the atomic and molecular gas.
For small regions to cool they must be disconnected from their surrounding for heat conduction.
This implies a closed magnetic structure, which is termed magnetic flux loop (tube).
Schematic drawing are in the upper panel of Figure \ref{fig:fig1} and in  Figure \ref{fig:fig4}
taken from Zoabi et al. (1996).
The loops might posses a more complicated structure such as depicted in the lower left loop
of the lower panel of Figure \ref{fig:fig1}.
Godon et al. (1998) discussed the role of magnetic flux loops in the cluster dynamo model
they proposed, and in the formation of cluster flares similar to solar flares
(see section \ref{subsubsec:alphaomega}).
Following works on the magnetic flux loop structure in CF clusters (Zoabi et al. 1996; Norman \& Meiksin 1996)
and on some similarities with stellar magnetic fields (Godon et et al. 1998), Kaastra et al. (2004)
extended the comparison with stellar magnetic fields to include the spectrum of magnetic flux loops.
They took the size of the loops to be $\sim 10 \kpc$.
(Kaastra et al. dropped the word ``flow'', and used the term ``cooling clusters''. This was a transition
phase to the term ``cool core clusters''.)
% FFFFFFFFFFFFFFFFFFFFFFFFFFFFF
% FFFFFFFFFFFFFFFFFFFFFFFFFFFFF

In the inner regions of the CF the magnetic fields are likely to be stronger, and the
dynamical effects of magnetic fields, such as tension, might become important,
as well as reconnection inside loops and between loops.
The magnetic tension might become important at the ends of the loops, where the magnetic
field lines are strongly bent (Zoabi et al. 1998).

% =========
\subsubsection{Heat conduction within flux loops}
\label{subsubsec:heat}
\begin{figure}
\begin{center}
\vskip11mm
\includegraphics[scale=0.9]{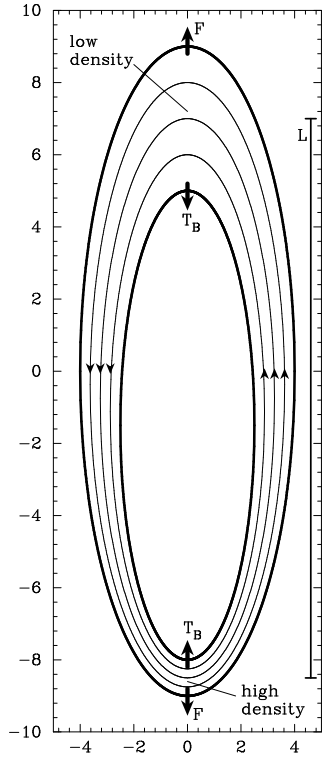}
\end{center}
\caption{A schematic structure of a magnetic flux loop taken from Soker (2004).
Units on axes are in kpc. With the jets and bubbles and their induced ICM motion,
somewhat smaller magnetic flux loops than those used by Soker (2004) are expected.
Thick arrows depicted forces: $T_B$ magnetic tension and $F$ buoyancy.
The thin ellipses are magnetic field lines, and $L$ is the length of the loop.
The lower segment is denser and cooler than its surroundings, while the upper segment
density is below that of its surroundings. After the lower segment cools to very low
temperatures, efficient heat conduction proceeds from the
upper segment to the lower one along magnetic field lines, hence reducing the
luminosity from X-ray emitting gas. }
\label{fig:fig5}
\end{figure}

% =========
The structure of magnetic flux loops allows heat conduction {\it within} the flux tube.
This has several consequences (Kaastra et al. 2004; Soker et al. 2004; Soker 2004).
The heat conduction will transfer heat from the hot X-ray emitting gas of the loop
to segments that are already below X-ray emitting temperatures. The cooler regions are dense
and efficiently radiate the energy at longer wavelengths.
This is schematically shown in Figure \ref{fig:fig5} taken from Soker (2004).
With the jets and bubbles and their induced ICM motion,
as discussed in previous sections here, smaller magnetic flux tubes
than those used by Soker (2004) are expected.
Short loops will ensure steeper temperature gradients,
and more efficient heat conduction. This process deserves more detailed study.
% FFFFFFFFFFFFFFFFFFFFFFFFFFFFF
% FFFFFFFFFFFFFFFFFFFFFFFFFFFFF

\begin{figure}
\begin{center}
\vskip11mm
\includegraphics[scale=0.6]{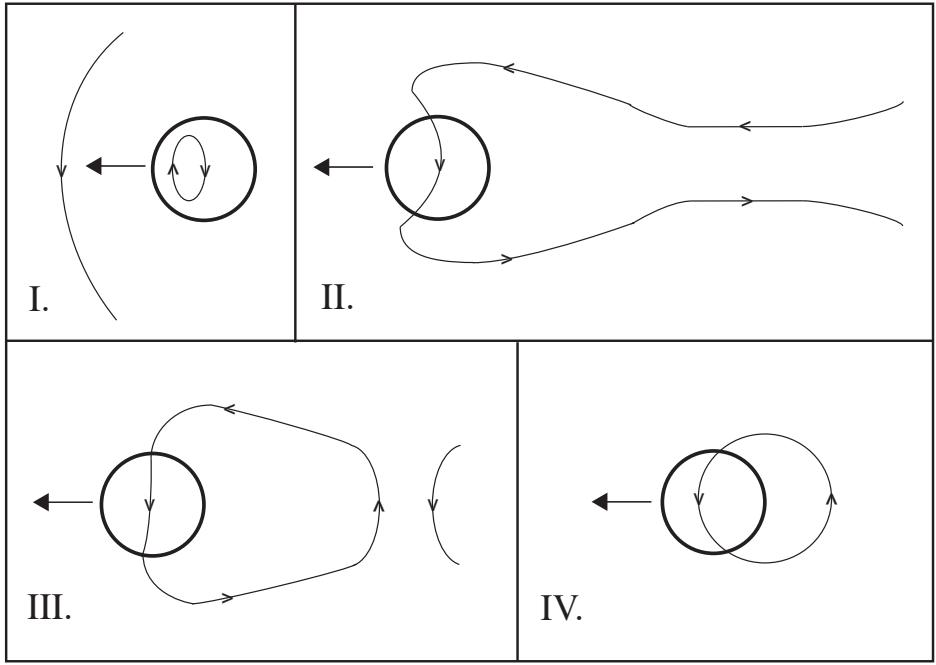}
\end{center}
\caption{A schematic evolution of a single intracluster magnetic field line encountering a
cold cloud.
The cold cloud moves to the left (panel I), such that the ambient field line collides with the magnetic field
inside the cloud. If the two fields are not aligned (they don't need to be exactly opposite),
reconnection occurs near the front of the cloud (the stagnation point), forming one field line
threading through the cloud and the ICM (panel II). Because of ICM material flowing away
from the region behind the cloud, the field line closes on itself behind the cloud
(panel II), leading to reconnection on the down-stream side, and the formation of a loop
connecting the cloud and ICM (panel III). Because only a limited volume of the hot ICM is
connected to the cold cloud, heat conduction cools the ICM, and the external pressure and
magnetic stress pull the conductively-cooling ICM toward the cloud (to the left in panel IV).
This process reduces the amount of X-ray radiation per cooling mass, as a large fraction
of the energy is emitted at longer wavelengths. (Taken from Soker et al. 2004.)}
\label{fig:fig6}
\end{figure}

Soker et al. (2004) have studied local heat conduction from cold ($T \sim 10^4 \K$) clouds
to their hot surrounding.
 The heat conduction is regulated by reconnection between the
magnetic field lines in the cold clouds and the field lines in the ICM.
A narrow conduction front is formed, which, despite the relatively
low temperature, allows efficient heat conduction from the hot ICM to the cold
clouds.
Narrow isothermal X-ray filaments are seen along the jets of M87 in the Virgo cluster
(Molendi 2002; Werner et al. 2010).
It seems that magnetic field lines were stretched along the jets of M87.
The reconnection between the field lines in cold clouds and those in the
ICM occurs only when the magnetic field in the ICM is strong enough.
This occurs only in the very inner regions of cooling flow clusters.
This process is depicted in Figure \ref{fig:fig6}.
Norman \& Meiksin (1996) have considered large scale magnetic flux loops, that reconnect
and by that allow global heat conduction. Their setting is different than the cases
discussed here, as they were aiming at globally suppressing cooling, while I study
only local heat conduction.
% FFFFFFFFFFFFFFFFFFFFFFFFFFFFF
% FFFFFFFFFFFFFFFFFFFFFFFFFFFFF

The processes shown in Figures \ref{fig:fig5} and \ref{fig:fig6} reduce the amount of emitted
X-ray per cooling mass, as more energy is emitted at longer wavelengths.
{{{ The heat conduction mentioned by Sparks et al. (2009) to explain emission from gas at a
temperature of $T\simeq 10^5 \K$ in M87 is an example for a local heat conduction process (see also
Sparks et al. 1989 for the emission gas in NGC 4696). }}}
Local heat conduction has similar effects to those of mixing hot and cold media in
CF clusters, as proposed by, e.g., Begelman \& Fabian (1990), Fabian et al. (2001, 2002) and
Bayer-Kim et al. (2002). In the mixing process the magnetic fields play a minor role as compared with
that in local heat conduction.

% =========
\subsubsection{An $\alpha-\omega$ dynamo in the very inner region}
\label{subsubsec:alphaomega}
% =========
Godon et al. (1998) discussed the possible operation of an $\alpha - \omega$ dynamo
in the inner $\sim 3 \kpc$ regions of CF clusters. The model was based in part
on the results of Soker \& Sarazin (1990).
It was assumed that the large scale inflow that was supposed to exist in
these models has some angular momentum. As the gas flows inward, the rotation
becomes more significant, but still negligible dynamically.
 Magnetic fields amplified by the inflow reconnect, and form a weak turbulence in the inner region.
The rotation and turbulence, Godon et al. (1998) speculated, can lead to the formation
of an $\alpha - \omega$ dynamo that further amplifies the magnetic fields.
As large massive inflow does not seem to exist, this model as was presented should
be modified.
Furthermore, the magnetic pressure inside $\sim 5-10 \kpc$ is limited to $\sim 10 \%$
of the thermal pressure (Churazov et al. 2008).

The presence of accretion disks around the central BH shows that some angular momentum exists
in the ICM. The angular momentum of the ICM is much lower than that assumed by Godon et al. (1998),
and might become significant only in the very inner regions of $<1 \kpc$.
However, the bubbles and jets activity make turbulence much stronger, as evident from
Figure \ref{fig:fig2}.
The possibility of an $\alpha-\omega$ dynamo, as well as amplification of magnetic fields
by turbulence alone,  in the very central region (of $r < 1\kpc$) in
CF clusters should be examined in more detail, despite that
the turbulent pressure seems to be only a few percent of the thermal pressure (David \& Nulsen 2008).

% =====================================================
\section{X-RAY DEFICIENT BUBBLES IN CF CLUSTERS}
\label{sec:bubbles}
% =====================================================

% ===================
\subsection{Properties of inflated bubbles}
\label{sec:bubblesG}
% ===================
I open this section with a summary of the main results that emerged from
our 2.5D hydrodynamical numerical simulations of bubble inflation.
When we consider inflated bubbles (rather than bubbles that are artificially inserted
to the ICM) some puzzles from the past are solved.
\newline
\textbf{(1)}
Fat bubbles can be inflated by slow massive wide (SMW) jets (Sternberg et al. 2007; Sternberg \& Soker 2008b, 2009a,b).
(Fat bubbles are wide bubbles attached, or almost attached, to the cluster center.)
Typical values are:  initial jet speed$\simeq 0.01-0.1c$;
mass loss rate into the two jets$\sim 1-100 M_\odot ~{\rm yr}^{-1}$;
a large half opening angle of $\alpha > 30^\circ$.
\newline
\textbf{(2)} Rapidly precession narrow jets have the same effect as wide jets
(Sternberg \& Soker 2008a, 2009a), and can also explain the formation of `fat' bubbles.
\newline
\textbf{(3)} Bubbles are known to be stable during the inflation phase
(Pizzolato \& Soker 2006). As discussed below in section \ref{sec:bubblesB},
when bubbles are inflated by appropriate jets, they stay intact for a much longer
time after the inflation phase ends (Sternberg \& Soker 2008b).
\newline
\textbf{(4)} Every bubble inflation episode excites multiple sound waves and shocks
(Sternberg \& Soker 2009a). Such sound waves are observed in
some CF clusters, most clearly in Perseus (Fabian et al. 2006).
\newline
\textbf{(5)} Mixing of very-hot bubble material with hot ICM gas can be a major channel to
heat the ICM (Sternberg \& Soker 2008b, 2009a; also Br\"uggen et al. 2009).
\newline
\textbf{(6)} The last two processes, as well as the shocks
that are found in the simulations, are effective not only in the directions of propagation
of the jets and bubbles, but also perpendicular to this symmetry axis.
It seems that these processes provide the energy channel from the jets to the ICM
in all relevant regions. Again, self-consistent inflation of bubbles by jets is required
to show that.
\newline
\textbf{(7)} Narrow very fast (even relativistic) jets can have negligible
energy content and still fill the bubbles with radio radiation (Soker et al. 2010).
I note that Cavagnolo et al. (2010) find that the energy in bubbles is much
larger than that radiated in the radio band.

% ===================
\subsection{No need for stabilizing magnetic fields inside bubbles}
\label{sec:bubblesB}
% ===================
I argue below that there is no need to assume the presence of magnetic fields
to stabilize bubbles in CF clusters.
(Magnetic fields do exist in bubbles, but it is not
clear if they even stabilize the bubbles.)

The estimated ages (Birzan et al. 2004) of most X-ray deficient bubbles in
CF clusters is about one order of magnitude larger than the characteristic
time of the Rayleigh-Taylor (RT) instability to destroy them under the assumption
of purely spherical bubbles rising in the undisturbed ICM (as I show, these are unrealistic assumptions).
This has prompted many authors (e.g., Br\"uggen \& Kaiser 2001; Kaiser et al. 2005;
Jones \& De Young 2005; Dong \& Stone 2009,) to invoke an ordered magnetic field at the
edge of the bubble, or to increase the viscosity (Reynolds et al. 2005; Dong \& Stone 2009;
Br\"uggen et al. 2009), to stabilize the bubble against the RT instability.
Indeed, numerical simulations of non-magnetic ``artificial bubble'' evolution show them
to be disrupted quite rapidly (e.g., Br\"uggen 2003; Br\"uggen \& Kaiser 2001; Jones \& De Young
2005; Pavlovski et al. 2008; Reynolds et al. 2005; Robinson et al. 2004;
Ruszkowski et al. 2008). Adding magnetic fields, it was claimed, make the bubbles more stable
(e.g., Jones \& De Young 2005; Robinson, K. et al. 2004; Ruszkowski et al. 2007).
However, it is not clear if magnetic fields can indeed supply the required stability
(Ruszkowski et al. 2007).
In the simulations cited above the bubbles were injected at off-center locations by
a prescribed numerical procedure. These are termed \emph{artificial bubbles}.
This is of course not the way bubbles are formed in clusters, and a more consistent
study is required.

The key to get the correct behavior is to consider the formation process of bubbles,
as has been done in recent years starting from jets (Sternberg et al. 2007;
Sternberg \& Soker 2008a,b, 2007a,b; Falceta-Goncalves et al. 2010; Morsony et al. 2010; Gaspari et al. 2010).
Sternberg \& Soker (2009a) inflated bubbles using slow, massive, wide (SMW)
jets with a wide opening angle, and followed their evolution as they rise
through the intra-cluster medium (ICM).
Such SMW outflows have been recently observed  to be blown by AGN (Moe et al. 2009,
and Dunn et al. 2010; see also claims for such outflows from X-ray observations
by  Tombesi et al. 2010, although they term these flows ultrafast jets
rather than slow jets).

When the realistic jet-inflation process of bubbles is considered there
are stabilizing processes.
\begin{enumerate}
\item During the inflation phase the bubble is RT stable because the bubble-ICM
interface is decelerating and expanding (Soker et al. 2002; Pizzolato \& Soker 2006).
\item The outward momentum of the bubble and the dense shell around it implies
that the low density bubble does not need to support a dense gas above it
during the outward motion to large distances (the dense shell is marked on
Figure \ref{fig:fig2}). This implies that the interface is RT stable.
\item It seems that the vortex inside the jet-inflated bubble (Figure \ref{fig:fig2})
stabilizes the sides of the bubble as it rises. Sternberg \& Soker saw no
sign of instabilities there in their many simulations, neither RT nor Kelvin-Helmholtz.
\end{enumerate}
 The result of these processes is that the bubbles can rise to large distances
from the cluster's center while still maintaining their general structure,
without the need to invoke stabilizing magnetic field.

{\bf Acknowledgement.} I thank Fulai Guo, Mark Voit, Larry David, Chris O'dea, and Brian McNamara for helpful comments.

\end{document}